\documentstyle[aaspp4]{article}
\begin{document}

\title{X-ray Nova XTE J1550-564: Optical Observations}

\author{Raj K. Jain\altaffilmark{1} and Charles
D. Bailyn} \affil{Department of Astronomy, Yale
University, P.O. Box 208101, New Haven, CT 06520-8101 \\
raj.jain@yale.edu, bailyn@astro.yale.edu}

\author{Jerome A. Orosz} \affil{Department of Astronomy \&
Astrophysics, The Pennsylvania State University, 525 Davey Laboratory,
University Park, PA 16802-6305 \\ orosz@astro.psu.edu}

\author{Ronald A. Remillard} \affil{Center for Space Research,
Massachusetts Institute of Technology, Cambridge, MA 02139-4307 \\
rr@space.mit.edu}

\and

\author{Jeffrey E. McClintock} \affil{Harvard-Smithsonian Center for
Astrophysics, 60 Garden Street, Cambridge, MA 02138-1516 \\
jem@cfa.harvard.edu}

\altaffiltext{1}{Also Department of Physics, Yale University,
P. O. Box 208120, New Haven, CT 06520-8120} 

\begin{abstract}

We report the identification of the optical counterpart of the X-ray
transient XTE~J1550-564 described in two companion papers by Sobczak
et al.\ (1999) and Remillard et al.\ (1999).  We find that the optical
source brightened by $\approx 4$ magnitudes over the quiescent
counterpart seen at $B\approx 22$ on a SERC survey plate, and then
decayed by $\approx 1.5$ magnitudes over the 7 week long observation
period.  There was an optical response to the large X-ray flare
described by Sobczak et al.\ (1999), but
it was much smaller and delayed by $\approx 1$
day.

\end{abstract}


\section{Introduction}

Soft X-ray transients, also called X-ray novae (XN), are mass
transferring binaries in which long periods of quiescence (when the
X-ray luminosity is $\le 10^{33}$ergs s$^{-1}$) are occasionally
interrupted by luminous X-ray and optical outbursts (Tanaka \&
Shibazaki 1996).  X-ray novae are unique objects since they provide
the most compelling evidence for the existence of stellar mass black
holes (Cowley 1992).  Using optical photometry and spectroscopy, eight
XN have been shown to contain a black hole, (van Paradijs \&
McClintock 1995; Bailyn et al.\ 1998; Orosz et al.\ 1998a) since the
mass of the primary exceeds the maximum stable limit of a neutron star
($\approx 3\,M_{\sun}$, Chitre \& Hartle 1976).

The soft X-ray transient XTE J1550-564 was discovered with the All Sky
Monitor (ASM ; Levine et al.\ 1996) on the {\em Rossi X-ray Timing
Explorer} (RXTE) on September 6, 1998 (Smith et al.\ 1998).  This
object became the brightest X-ray nova yet observed by RXTE (Remillard
et al.\ 1998b).  A high frequency QPO at $\approx 185$ Hz has been
observed on two well separated occasions (McClintock et al.\ 1998;
Remillard et al.\ 1999).  Although the true nature of this object has
yet to be confirmed, XTE J1550-564 is likely to be a black hole based
on its characteristic soft X-ray spectrum and the hard power law tail
(Sobczak et al.\ 1999), and high frequency QPOs (Remillard et al.\
1999).

Shortly after the X-ray discovery the optical counterpart was
identified within the RXTE error box (Orosz, Bailyn \& Jain 1998).  We
present optical light curves obtained during the outburst of XTE
J1550-564 as part of a multi-wavelength campaign.  
The spectral analysis of the RXTE PCA data as well as the ASM
light curve and a 
timing study based on the same RXTE observations are presented in companion
papers (Sobczak et al.\ 1999 and Remillard et al.\ 1999; hereafter paper I
and paper II, respectively). 

Although much can be learned about XN by studying the X-ray data
alone, simultaneous optical observations can provide tighter
constraints for various accretion disk models.  For example, optical,
UV and X-ray data obtained during quiescence of several BHXN have been
used to demonstrate the successful application of the advection
dominated accretion flow (ADAF) model, whereas thermal emission from a
thin disk model is inconsistent with these observations (Narayan et
al.\ 1996, 1997a,b).  Furthermore, the six day time delay between the
optical and X-ray outbursts of GRO J1655-40 observed in April 1996
(Orosz et al.\ 1997) has also been successfully modeled by an
accretion flow consisting of a cold outer disk and a hot inner ADAF
region (Hameury et al.\ 1997).  The extensive X-ray and optical
coverage of the outburst of XTE~J1550-564 provides further
opportunities to test accretion disk models and ADAF models in
particular.  We report below our optical observations, data
reductions, and results.

\section{Observations and Reductions} 

We obtained photometry using the Yale 1m telescope at CTIO, which is
currently operated by the YALO (Yale, AURA, Lisbon, Ohio State)
consortium.  This telescope is ideally suited for observing X-ray
transients and other objects which require continuous long-term
monitoring.  Data are taken every clear night by two permanent staff
observers, and observations are requested by a queue which can be
changed quickly in response to discoveries.  The data reported here were
acquired using the ANDICAM optical/IR camera which
contains a TEK $2048\times 2048$ CCD with $10.2 \times 10.2$
arcmin$^2$ field of view with a scale of 0.3 arcsec
pixel$^{-1}$. The IR array was not available at the time of our
observations.

On September 8.99 (UT), we obtained images of two fields in response
to the announcement of the initial detection of XTE J1550-564 by RXTE
(Smith et al.\ 1998).  For each field a 60 second and a 300 second
exposure was obtained in both $V$ and $I$.  The 300 second $V$ band
images were compared to images of the same regions extracted from the
Digitized Sky Survey (DSS) CD ROM set (Sturch et al.\ 1993).  We
identified a $V\approx 16$ star as the optical counterpart, since it
appeared in all of the CCD images we obtained, but not in the DSS
image (Orosz, Bailyn, \& Jain 1998 --- see Figure~1). 
There were several HST guide stars in the CCD image, which allowed
us to determine the
J2000 coordinates of $\alpha=15^{\rm h}50^{\rm
m} 58\fs 78$, $\delta=-56^\circ 28^{\prime} 35\farcs 0$, with errors of
$\lesssim 2^{\prime\prime}$.  The quoted
error corresponds to the maximum
systematic error present in the HST Guide Star Catalog (Russell et al.\ 
1990).
A variable radio source 
detected on September 9 at a position consistent with that of the optical
transient 
(Campbell-Wilson et al.\ 1998).
Finally, spectroscopic confirmation
came on September 16, when Castro-Tirado et al.\ (1998) showed that
the optical variable had emission lines of H, He II, and N III,
typical of X-ray transients in outburst.

We observed the source on all nights for which weather and
instrumentation permitted between September 8.99 to October 26.9,
1998, when the object was no longer observable in the night sky (see
Figure~2).  The exposure times were 120-300 seconds for $V$ and $I$,
300-600 seconds for $B$.  The seeing varied from night to night
ranging from 1.3 to 3 arcseconds, with a typical value of 1.7
arcseconds.  A time series of the optical data was obtained using the
IRAF versions of DAOPHOT and ALLSTAR and the stand-alone code
DAOMASTER (Stetson 1987; Stetson, Davis, \& Crabtree 1991; Stetson
1992a, 1992b). The DAOPHOT instrumental magnitudes were calibrated to
the standard scales using standard stars from the list of Landolt
(1992).

\section{The Quiescent Optical Counterpart}
 
The Royal Observatory Edinburgh (ROE) maintains a large archive of
photographic plates of Southern sky fields taken with the UK Schmidt
telescope located at the Anglo-Australian Observatory.  There are ten
plates on which the XTE J1550-564 field was well-centered and the
exposure times were fairly long.  Sue Tritton of the ROE kindly
examined these ten plates and photographed the $\approx 4\times 4$
arcminute regions surrounding the positions of XTE J1550-564.  The
best quality plate is No.\ J2977 from March 20, 1977, which is the
atlas plate for the SERC J survey.  A print of this plate was scanned
using the Yale PDS-microdensitometer at a resolution of 0.3 arcseconds/pixel.
There is a faint star close to the
position of XTE J1550-564 (see Figure 1).   We used the pixel coordinates
of several bright comparison stars to determine the coordinate transformation
between the CCD image and the scanned image.  Based on this transformation,
we find that the faint star in the scanned image is within 0.5 arcseconds
of the position of XTE J1550-564.  
Hence this
star is most likely the quiescent optical counterpart, although we can
not completely rule out the possibility that it is an unrelated field
star.  We performed aperture photometry of this faint star and several
comparison stars in the scanned image and determined a $B$ magnitude
of $B=22.0\pm 0.5$ for the faint counterpart. 
This error represents the scatter
in the differences between the calibrated magnitudes of the comparison stars
and the magnitudes obtained from the
scanned photographic image.  
The quiescent
counterpart is not visible on any of the the remaining nine plates,
all of which have limiting magnitudes of $\lesssim 21$.

The amplitude of the outburst thus appears to be $\approx 4$
magnitudes, although this could be larger if the star appearing on the
SERC J plate is not the actual quiescent counterpart.  The outburst
amplitude is intermediate between the relatively small optical
outbursts seen in X-ray novae with early type (F and A) companions
like GRO~J1655-40 and 4U1543-47 (Bailyn et al.\ 1995; Orosz et al.\
1998a) and the much larger outbursts seen in systems with later
spectral types (van Paradijs \& McClintock 1995). If this pattern
holds, one may expect the secondary of XTE~J1550-56 to be a
main-sequence G star or perhaps an evolved giant.  For main sequence
companions, Shahbaz \& Kuulkers (1998) propose an empirical formula
relating the orbital period to the outburst magnitude, which yields $P
\approx 23$ hours for this case.

\placefigure{fig1}

\section{The Outburst Light Curve}

The optical magnitude of XTE J1550-564 varied much less during our
observing period than the X-ray flux (see Figure 2).  During the span
of 49 days that we have data for XTE J1550-564, the optical brightness
in $B$, $V$ and $I$ dropped by $\approx 1.5$ magnitudes and the daily
fluctuations were less than $\approx 0.15$ magnitude between any two
adjacent nights.  In general, the $B$, $V$, and $I$ magnitudes
decayed steadily with the exception of an optical flare near
September 21, which occurred approximately one day after the X-ray
flare.  There is also a seven day plateau lasting from approximately
October 15 to October 21, during which the $I$ magnitude fluctuated by
less than 0.03 magnitudes (the $B$ and $V$ magnitudes also fluctuated
much less during this period compared to the previous seven days).

The general features of the optical and X-ray light curves can be
compared with other LMXBs using the classifications by Chen, Shrader
\& Livio (1997).  Based on their classification, the exponentially
decaying optical light curve of XTE J1550-564 with an e-folding time
of $\approx 30$ days can be described as a possible FRED (light curves
that have either a Fast-Rise or an Exponential Decay).
In fact, most optical light curves of LMXBs are best described as
possible FREDs, although the average e-folding time of 67.6 days
(Chen, Shrader, \& Livio 1997) is longer than what we find for
XTE J1550-564. Curiously, the optical decay is not correlated
with the ASM X-ray light curve, which remains fairly constant
after the flare.

Before we can compare properties such as the average intrinsic color
and ratio of the X-ray to optical flux of XTE J1550-564 to those of
other LMXBs, we must correct for interstellar reddening.  The
reddening can be estimated from the average expected values of $N_H$
which can be derived from the HI map by Dicky \& Lockman (1990), in
this case by using the FTOOLS routine nh.  We obtained $N_H \approx 9
\times 10^{21}$ cm$^{-2}$, which yields an estimate of $A_V=5.0$
assuming the relation between $N_H$ and $A_V$ of Predehl \& 
Schmitt (1995).  The Dicky \& Lockman (1990) map represents the
column density to infinity, and thus might be an overestimate for
a galactic source embedded in the plane.  However, the tentative
distance of 6 kpc suggested in paper I places the source 200pc below
the plane, well out of the galactic dust layer.  We note that the
X-ray spectral analysis yields values of $N_H$ about twice as
large as is suggested by the HI maps --- this may indicate
self-absorption in the source.

Using the conventional
relationship $A_V = 3.1 \times E(B-V)$ (Savage \& Mathis 1979), we
find $E(B-V)=1.6$.  This implies an intrinsic color of $(B-V)_0
= -0.25\pm 0.04$ right after the flare, consistent
with the average value of $-0.09 \pm 0.14$ for a sample of LMXBs
obtained by van Paradijs \& McClintock (1995).  Note that the
much larger reddening implied by the X-ray column density in the
absence of self-absorption results
in an implausibly blue intrinsic color for XTE J1550-564.  Using
$A_V=5.0$, we find an optical to X-ray flux ratio of $\approx 450$
during the flare, comparable to the average value of 500 found
by van Paradijs \& McClintock (1995).  This estimate assumes
a flat optical spectrum between $3000\AA $ and $7000\AA $, and
a 2-20 keV X-ray flux derived from the spectral decomposition 
described in paper I.

The optical response to the large X-ray flare which occurred near
Sept. 21 (see paper I) was quite muted, with an
amplitude of $\le 0.2$
magnitudes. The (V-I) color increased during the flare, meaning the
optical outburst was redder than during the decay (see Figure 2). On
the other hand, the ASM hardness ratio (HR2, defined as the ratio of
ASM count rates between 5 to 12 keV band and 3 to 5 keV bands)
increased, meaning the X-ray light curve was ``bluer'' during the
flare (see paper I for details).  Although detailed
spectral information about the optical flare is unavailable, the color
and hardness ratios indicate a change in the spectrum across a wide
range of wavelengths during the flare.

\placefigure{fig2}

The optical response to the flare is delayed by about a day relative
to the X-rays.  To quantify this delay,
we parameterized the light curves as described below.
The fits are not perfect, and somewhat different results can be
obtained with different fitting schemes.  However the qualitative
relations between the different light curves persist no matter how the
data are described.  We fit the ASM flare with a Lorentzian with a
centroid of $51,076.2 \pm 0.1$ and a FWHM of $\approx 1.5$ days (for
convenience we express time in days in the units of
MJD=JD-2,400,000.5, where MJD 51,075 is September 19, 1998).  We
estimated the centroid and FWHM of the optical flare using a Gaussian
to fit the flare and a linear component to fit the decay.  We
determined the centroids for the $B$, $V$, and $I$ bands to be
$51,077.3 \pm 0.2$, $51,077.05 \pm 0.02$, $51,077.05 \pm 0.03$,
respectively, all of which occur approximately one day later than the
X-ray peak as noted above.  There is no significant offset between the
times of the peaks of the optical flare in the $B$, $V$, and $I$ light
curves.  In contrast, the duration of the optical event varied
strongly with filter.  One self-consistent solution yields the
following values of FWHM for $B$, $V$, and $I$ respectively: $2.1 \pm
0.5$, $1.6 \pm 0.1$, and $1.6 \pm 0.05$ days.
The relationship between the X-ray and optical flares is in dramatic
contrast to the onset of the April 1996 outburst of GRO~J1655-40,
described by Orosz et al.\ (1997), in which the optical event
preceded the X-rays by six days.  That event was interpreted
(Hameury et al. 1997)
as an ``outside-in'' instability in the accretion disk.  The flare
in XTE~J1550-564, on the other hand, appears to have begun deep in
the accretion flow, and subsequently propagated outwards.

\section {Summary}

We have identified the optical counterpart of XTE~J1550-564, and
analyzed the B, V and I light curves from
September 8.99 to October 26.9, 1998.  We find that the
X-ray and optical light curves are poorly correlated.  The large
X-ray flare was followed a day later by a small ($0.2$ mag)
increase in
the optical brightness.  The tentative identification of a 
quiescent counterpart in sky survey images suggests that it will
be possible to measure the mass function of the object after
it has returned to quiescence.

\acknowledgements

We thank the two YALO observers, David Gonzalez Huerta and Juan
Espinoza, for providing data in a timely manner.  We also thank Sue
Tritton of the ROE for her help with the archival plates and John Lee,
Terry Girard, and Imants Platais for assistance with the Yale PDS
scanner. We would like to thank Suzanne Tourtellotte and Elene Terry
for their assistance with data reduction.  Financial support for this
work was provided by the National Science Foundation through grant
AST-9730774.

\clearpage

\figcaption[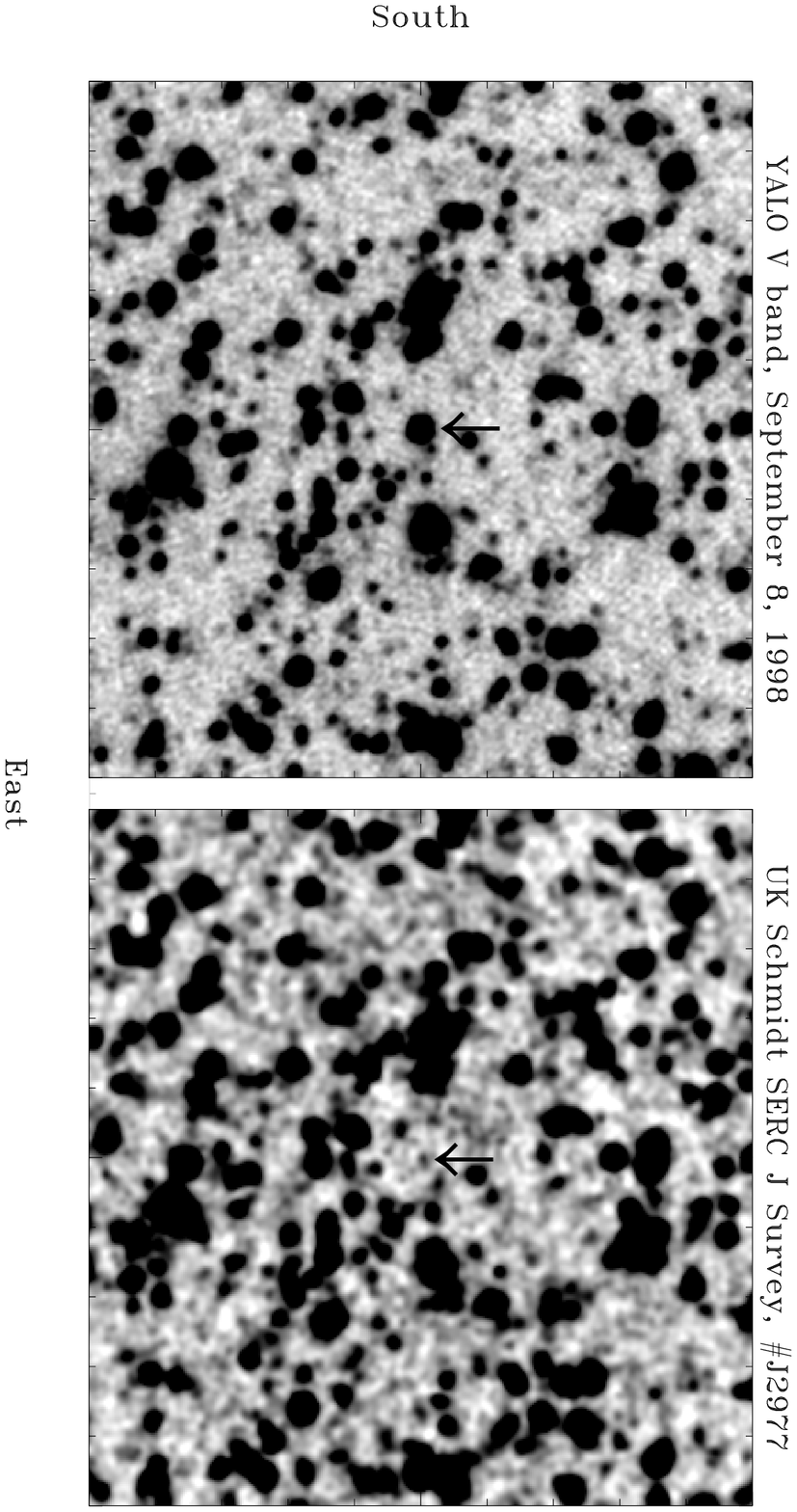]{left: The YALO 1m telescope image (2 by 2
arcminutes).  The optical counterpart is marked with the
arrow.  Right:
A scanned image of the 1.22m UK Schmidt SERC J Survey plate J2977
showing the same region of the sky.  The arrow points to the optical
counterpart in quiescence.  Both images have been smoothed by
convolution with a Gaussian (width = 1.5 pixels) and stretched by a
similar amount to bring out faint objects. There is a slight mismatch
between the effective bandpasses of the two images. 
\label{fig1}}

\figcaption[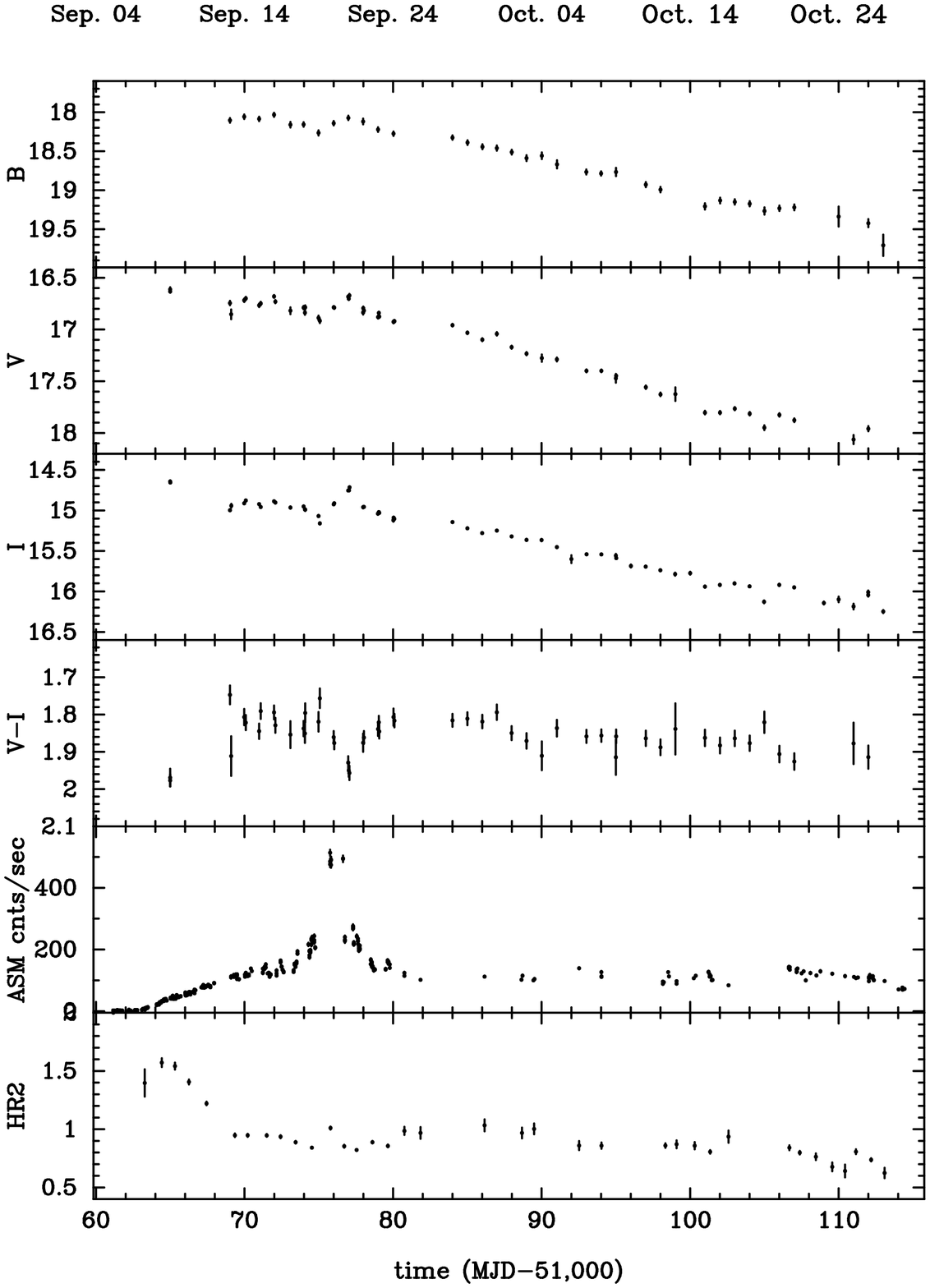]{ from top to bottom: The B, V, I, and V-I
light curves.  RXTE All Sky Monitor light curve (note the bright flare
near day 51077), and the RXTE HR2, defined as the ratio of ASM count
rates between 5 to 12 keV and 3 to 5 keV. MJD is defined to be JD -
2,400,000.5.
\label{fig2}}

\clearpage

\plotone{ffind.ps}

\plotone{flc.ps}

\end{document}